\documentclass[a4paper,10pt]{article}

\title{Shape Dynamics in 2+1 Dimensions}
\author{\bf Timothy Budd\footnote{\href{mailto:t.g.budd@uu.nl}{t.g.budd@uu.nl}}
\\\it Institute for Theoretical Physics, Utrecht University\\\it Postbus 80195, 3508 TD, Utrecht, The Netherlands\bigskip\\ \bf Tim Koslowski\footnote{\href{mailto:tkoslowski@perimeterinstitute.ca}{tkoslowski@perimeterinstitute.ca}}
\\\it Perimeter Institute for Theoretical Physics\\\it 31 Caroline Street, Waterloo, Ontario N2L 2Y5, Canada}

\usepackage{amssymb}
\usepackage{amsmath}
\usepackage[usenames,dvipsnames]{color}
\usepackage{hyperref}
\usepackage[utf8x]{inputenc}
\usepackage[left=1in,right=1in,top=1in,bottom=1in]{geometry}                  % For good copy.

\hypersetup{
    colorlinks=true,         % false: boxed links; true: colored links, false is default
    linkcolor=blue,          % color of internal links, red is default
    citecolor=red,        % color of links to bibliography, 'green' is default
    urlcolor=Violet             % color of external links, cyan is default
}

\let\oldmarginpar\marginpar
\renewcommand\marginpar[1]{\oldmarginpar{\color{red}\raggedright\scriptsize #1}}

\begin{document}

\maketitle

\begin{abstract}
  Shape Dynamics is a formulation of General Relativity where refoliation invariance is traded for local spatial conformal invariance. In this paper we explicitly construct Shape Dynamics for a torus universe in 2+1 dimensions through a linking gauge theory that ensures dynamical equivalence with General Relativity. The Hamiltonian we obtain is formally a reduced phase space Hamiltonian. The construction of the Shape Dynamics Hamiltonian on higher genus surfaces is not explicitly possible, but we give an explicit expansion of the Shape Dynamics Hamiltonian for large CMC volume. The fact that all local constraints are linear in momenta allows us to quantize these explicitly, and the quantization problem for Shape Dynamics turns out to be equivalent to reduced phase space quantization. We consider the large CMC-volume asymptotics of conformal transformations of the wave function. We then use the similarity of Shape Dynamics on the 2-torus with the explicitly constructible strong gravity (BKL) Shape Dynamics Hamiltonian in higher dimensions to suggest a quantization strategy for Shape Dynamics.
\end{abstract}

\section{Introduction}\label{sec:Introduction}

Shape Dynamics is constructed as reformulation of General Relativity in which spacetime refoliation invariance is traded for spatial conformal invariance that preserves the total spatial volume \cite{Barbour:2011dn,Gomes:2010fh,Gomes:2011zi} using of a linking gauge theory. The development of Shape Dynamics was inspired by Dirac's work \cite{Dirac:1958jc} on CMC (constant mean extrinsic curvature) gauge, York's method for solving the initial value problem \cite{York:1973ia,York:york_method_prl,Niall_73} and Machian ideas developed by Barbour and collaborators \cite{Anderson:2004bq,Barbour:2010xk}. The explicit construction of the Shape Dynamics Hamiltonian however requires the general solution of a partial differential equation, which is equivalent to partially solving the initial value problem of General Relativity using York's method. This complication obstructs many straightforward investigations. To learn about Shape Dynamics it is therefore very valuable to consider exactly solvable nontrivial gravitational models. This provides the main motivation for this paper: We consider a nontrivial model in which Shape Dynamics can be constructed explicitly allowing us to study its generic features.

Probably the best known example of a nontrivial exactly solvable gravitational system is pure gravity on the torus in 2+1 dimensions \cite{Martinec:1984fs,Moncrief:1989dx,Carlip:book,Carlip:2004ba}. The technical reason for the simplifications in this model is two-fold: First, one is able to solve the initial value problem of ADM gravity explicitly on the 2-torus. This is important for the construction of classical Shape Dynamics and occurs only on the 2-torus and 2-sphere;\footnote{The sphere is a degenerate case, since it admits only one canonical pair of degrees of freedom (the volume and the mean extrinsic curvature). This admits only the de Sitter solution which contains no interesting dynamics.} pure gravity on higher genus surfaces is more intricate since we lack methods to solve the initial value problem in general. Second, the reduced phase space (after solving for initial data) is finite dimensional, which is a generic feature of pure gravity in 2+1 dimensions. This is important for quantization, because a finite dimensional system admits generic quantum theories while nontrivial quantum systems with infinitely many degrees of freedom are sparse.

The plan for this paper is as follows: We start with the explicit construction of pure Shape Dynamics on the 2-torus in section \ref{sec:Equivalence} and show its equivalence with General Relativity using the method of linking gauge theories. The trading of refoliation invariance for local spatial conformal invariance turns all local constraints into phase space functions that are linear in the momenta, while the remaining Shape Dynamics Hamiltonian turns out to formally coincide with the reduced phase space Hamiltonian, which at large CMC-volume becomes the conformal constraint that changes the total volume. In section \ref{sec:HigherGenus} we consider Shape Dynamics on a higher genus surface, which does not admit straightforward solutions to the initial value problem. We attack this problem by constructing a perturbation expansion and we recover a fully conformal theory in the large volume regime. We find that the generically nonlocal Hamiltonian becomes the integral over a local density at large volume and turns again into the conformal constraint that changes the total volume. We then use the classical results to quantize Shape Dynamics on the torus in section \ref{sec:Quantize}. Due to linearity of the local constraints, one can implement them at the quantum level and thus quantize the analogue of the Wheeler--DeWitt equation, i.e. infinitely many degrees of freedom. In an excursion we discuss the large-CMC-volume asymptotics of Shape Dynamics wave functions, which resembles aspects of a quantum gravity/CFT correspondence. Before we conclude in section \ref{sec:Conclusions} we discuss which results of this paper could generalize to higher dimensions and which are specific to 2+1 dimensions in section \ref{sec:HigherDimensions}.

\section{Equivalence of General Relativity and Shape Dynamics}\label{sec:Equivalence}

In this section we establish the equivalence between General Relativity and Shape Dynamics on the 2+1 dimensional torus universe by explicitly constructing the linking theory relating the two. For simplicity we assume a positive cosmological constant $\Lambda$. We start with the general construction of the linking theory before focusing on the torus, which allows us, other than in higher dimensions or even on a higher genus surface, to explicitly work out Shape Dynamics.

A convenient way to construct the linking theory is to follow the ``best matching procedure'' \cite{Barbour:2011dn} starting with the ADM Hamiltonian on the usual ADM phase space $\Gamma_{ADM}$, expressed in terms of the metric $g_{ab}(x)$ and its canonically conjugate momentum density $\pi^{ab}(x)$:
\begin{equation}\label{equ:ADMconstraints}
  \begin{array}{rcl}
    H&=&S(N)+ H(\xi),\\
    S(N)&=&\int d^2x \,N\left( \frac{1}{\sqrt{|g|}}\pi^{ab}G_{abcd}\pi^{cd}-\sqrt{|g|}\left(R-2\Lambda\right)\right),\\
    H(\xi)&=&\int d^2x\, \pi^{ab}\mathcal L_\xi g_{ab}.
  \end{array}
\end{equation}
Here $G_{abcd}=g_{ac}g_{bd}-g_{ab}g_{cd}$ is the 2-dimensional super-metric, and $S$ and $H$ denote the ADM scalar and diffeomorphism constraints. To best match with respect to conformal transformations that preserve the total volume we consider the ADM phase space as a subspace of a larger phase space $\Gamma_{ext}=\Gamma_{ADM}\times \Gamma_\phi$, where $\Gamma_\phi$ is the phase space of a scalar field $\phi(x)$, whose canonically conjugate momentum density is denoted by $\pi_\phi(x)$. The phase space functions on $\Gamma_{ADM}$ are naturally identified with those phase space functions on $\Gamma_{ext}$ that are independent of $(\phi,\pi_\phi)$. We can thus recover usual ADM gravity in this larger system by introducing an additional first class constraint
\begin{equation}\label{equ:additionalConstraint}
  Q(x):=\pi_\phi(x)\approx 0
\end{equation}
and add it smeared with a Lagrange multiplier $\rho(x)$ to the ADM Hamiltonian, which is now $H=S(N)+H(\xi)+Q(\rho)$. Let us now consider a canonical transformation from $(g_{ab},\pi^{ab},\phi,\pi_\phi)$ to $(G_{ab},\Pi^{ab},\Phi,\Pi_\phi)$ generated by the generating functional
\begin{equation}\label{equ:generatingFunctional}
 F=\int d^2x\left(g_{ab}(x)e^{2\hat\phi(x)}\Pi^{ab}(x)+\phi(x)\Pi_\phi(x)\right).
\end{equation}
Here $\hat{\phi}$ is defined in terms of $\phi$ by subtracting a spatial average, that has a non-trivial dependence on the metric,
\begin{equation}
  \hat \phi (x):=\phi(x)-\frac 1 2 \ln \left\langle e^{2 \phi} \right\rangle_g,
\end{equation}
where we use the shorthands $\langle f\rangle_g=V_g^{-1}\int d^2x \sqrt{|g|} f$ and $V_g=\int d^2x \sqrt{|g|}$. Notice that we constructed $\hat \phi$ such that the conformal factor $e^{2\hat \phi}$ preserves the total volume. The canonical transformation of the elementary variables can be worked out explicitly:
\begin{equation}
 \begin{array}{rcl}
   G_{ab}(x)&=&e^{2\hat \phi(x)}g_{ab}(x),\\
   \Pi^{ab}(x)&=&e^{-2\hat \phi(x)}\left(\pi^{ab}(x)-\frac 1 2\sqrt{|g|}(x)g^{ab}(x)\langle\pi\rangle\left(1-e^{2\hat\phi(x)}\right)\right),\\
   \Phi(x)&=&\phi(x),\\
   \Pi_\phi(x)&=&\pi_\phi(x)-2\left(\pi(x)-\langle\pi\rangle\sqrt{|g|}(x)\right),
 \end{array}
\end{equation}
using shorthand notation $\pi(x)=\pi^{ab}(x)g_{ab}(x)$ and $\langle\pi\rangle=V^{-1}\int d^2x\,\pi(x)$. This transformation leads us to the constraints of the linking theory:
\begin{equation}\label{equ:linkingConstraints}
  \begin{array}{rcl}
    H&=&S(N)+H(\xi)+Q(\rho),\\
    S(N)&=&\int d^2x N\Bigl[\frac{e^{-2\hat\phi}}{\sqrt{|g|}}\left(\pi^{ab}G_{abcd}\pi^{cd}-\frac 1 2 \left(\pi-\langle\pi\rangle(1-e^{6\hat\phi})\sqrt{|g|}\right)^2+\frac 1 2 \pi^2\right)\\
    & &-\sqrt{|g|}\left(R[g]-2\Delta \hat \phi-2\Lambda e^{2\hat\phi}\right)\Bigr],\\
    H(\xi)&=&\int d^2x e^{-2\hat \phi(x)}\left(\pi^{ab}(x)-\frac 1 2\sqrt{|g|}(x)g^{ab}(x)\langle\pi\rangle\left(1-e^{2\hat\phi(x)}\right)\right)\left(\mathcal{L}_\xi e^{2\hat \phi(x)}g\right)_{ab}(x), \\
    Q(\rho)&=&\int d^2 x \rho(x)\left(\pi_\phi(x)-2\left(\pi(x)-\langle\pi\rangle\sqrt{|g|}(x)\right)\right).
  \end{array}
\end{equation}
One can check that after integrating by parts and using $Q=0$ the constraint $H(\xi)$ turns into the usual form of the diffeomorphism constraint $H(\xi)=\int d^2x\left(\pi^{ab}\mathcal{L}_\xi g_{ab}+\pi_\phi\mathcal{L}_\xi\phi\right)$. We will use this form of the constraint below. The linking theory thus contains the usual diffeomorphism constraint, a conformal constraint that preserves the total 2-volume and a scalar constraint that arises as a modification of the ADM refoliation constraint.

\subsection{Linking Theory on the Torus}

We will now exploit some special properties of two dimensional metrics on the torus to simplify the constraints (\ref{equ:linkingConstraints}). First of all, it is a well-known fact that all metrics on the torus are conformally flat. The space of flat metrics modulo diffeomorphisms is finite dimensional and admits a convenient parametrization of the space of metrics on the torus. We follow \cite{Carlip:2004ba,Carlip:book} where possible.

To make this more explicit let us fix a global chart on the 2-torus $\mathbb T^2$, which allows us to uniquely identify any point $x\in\mathbb T^2$ with its coordinates $(x^1,x^2)\in[0,1)^2$. In these coordinates we can decompose an arbitrary metric $g_{ab}$ as
\begin{equation}\label{equ:metricDecomposition}
g_{ab}(x)=e^{2\lambda(x)}\left(f^* \bar g\right)_{ab}(x),
\end{equation}
where $\lambda$ is a conformal factor, $f$ is a (small) diffeomorphism and $\bar{g}$ a flat reference metric. We can make this decomposition unique by requiring $\bar{g}$ to be of the form
\begin{equation}\label{equ:standardMetricTorus}
  \bar g=\frac{1}{\tau_2}\left( dx^1\otimes dx^1+\tau_1 (dx^1\otimes dx^2+dx^2\otimes dx^1)+(\tau_1^2+\tau_2^2)dx^2\otimes dx^2 \right)
\end{equation}
where $\tau=(\tau_1,\tau_2)$ denote the Teichm\"uller parameters. There is a slight redundancy left in the decomposition having to do with the fact that $f$ is only determined up to an isometry of $\bar{g}$, i.e. up to translations in $x^1$ and $x^2$. If we require $f$ to leave $(0,0)$ invariant, we really obtain a one to one map between metrics on the torus and the data $(\lambda,f,\tau)$.

We can explicitly decompose the momentum density\footnote{All indices here are raised with the reference metric $\bar g$}
\begin{equation}\label{equ:momentumDecomposition}
 \pi^{ab}=e^{-2\lambda}\left(p^{ab}+\frac 1 2 \pi \bar g^{ab}+\sqrt{|\bar g|}\left(\bar D^a Y^b+\bar D^b Y^a-\bar g^{ab}\bar D_c Y^c\right)\right),
\end{equation}
in terms of a trace $\pi$, a vector field $Y$ and a transverse traceless tensor density (w.r.t. $\bar{g}$), which we can explicitly parametrize by
\begin{equation}\label{equ:standardMomentumTorus}
 p=\frac 1 2\left(\left((\tau_1^2+\tau_2^2)p_2-2\tau_1\tau_2p_1\right)\partial_1\otimes\partial_1+\left(\tau_2p_1-\tau_1p_2\right)\left(\partial_1\otimes\partial_2+\partial_2\otimes\partial_1\right)+p_2\,\partial_2\otimes\partial_2\right).
\end{equation}
This decomposition is such that $\pi$ is conjugate to $2\lambda$ and $p_i$ is conjugate to $\tau_i$.

Writing the linking theory constraints (\ref{equ:linkingConstraints}) in terms of $\lambda,f,\tau,\pi,Y$ and $p$ we get
\begin{equation}\label{equ:torusConstraints}
 \begin{array}{rcl}
   S(N)&=&\int d^2 x N\left[\frac{e^{-2(\hat\phi+\lambda)}}{\sqrt{|\bar g|}}\left((p^{ab}+(PY^{ab})\sqrt{|\bar g|})\bar g_{ac}\bar g_{bd}(p^{cd}+(PY^{cd})\sqrt{|\bar g|})\right.\right.\\
   &&\left.\left.-\frac 1 2 (\pi-\langle\pi\rangle\sqrt{|\bar g|}(1-e^{2(\hat \phi+\lambda)}))^2\right)-\sqrt{|\bar g|}\left(k-2\bar \Delta(\hat \phi+\lambda)-2\Lambda e^{2(\hat \phi+\lambda)}\right)\right]\\
   H(\xi)&=&\int d^2 x \xi^a\left(\sqrt{|\bar g|}\bar \Delta Y_a+\frac 1 2 e^{2\lambda}\bar D_a(e^{-2\lambda}\pi)+\pi_\phi \phi_{,a}\right)\\
   Q(\rho)&=&\int d^2 x \rho(x)\left(\pi_\phi(x)-2\left(\pi(x)-\langle\pi\rangle\sqrt{|\bar g|}(x)e^{2\lambda(x)}\right)\right),
 \end{array}
\end{equation}
where we used the shorthand $PY_{ab}=\bar D_a Y_b+\bar D_b Y_a-\bar g_{ab}\bar g^{cd} \bar D_c Y_d$.

To complete the definition of a linking theory, we specify two sets of gauge-fixing conditions,
\begin{equation}
  \phi(x)=0\,\,\,\textrm{ for GR and }\,\,\,\pi_\phi(x)=0\,\,\,\textrm{ for SD,}
\end{equation}
which we will now use to reconstruct General Relativity and Shape Dynamics respectively.

\subsection{Recovering General Relativity}

To recover standard ADM gravity on the torus let us impose the gauge-fixing condition $\phi(x)=0$ to the linking theory. To perform the phase space reduction from the extended phase space to the ADM phase space, we need to fix the Lagrange multipliers such that the gauge-fixing is propagated. Since the momentum density $\pi_\phi$ occurs only in the constraints $Q$, we have to solve
\begin{equation}
  0=\{Q(\rho),\phi(x)\}=\rho(x)
\end{equation}
for the Lagrange-multiplier $\rho$, which implies $\rho=0$, hence the constraints $Q(x)$ are gauge-fixed and in fact drop out of the Hamiltonian, which becomes independent of $\pi_\phi$. Hence, one can perform the phase space reduction by setting $\phi\equiv 0$, $\rho\equiv 0$ and $\pi_\phi$ arbitrary\footnote{Had the constraints $Q(\rho)$ not dropped out after gauge fixing $\rho$, we would have had to solve $Q\equiv 0$ for $\pi_\phi$ to complete the phase space reduction.} in equation (\ref{equ:linkingConstraints}). The Hamiltonian on the ADM phase space thus reads
\begin{equation}
  \begin{array}{rcl}
    H&=&S(N)+H(\xi),
  \end{array}
\end{equation}
where $S(N)$ and $H(\xi)$ are precisely the scalar- resp. diffeomorphism constraint of General Relativity in the ADM formulation as given in equation (\ref{equ:ADMconstraints}). We note that we explicitly retained refoliation invariance.

\subsection{Recovering Shape Dynamics}\label{sec:SDon2torus}

To recover Shape Dynamics we employ the gauge-fixing condition $\pi_\phi(x)=0$. We will see that the decomposed form (\ref{equ:torusConstraints}) of the constraints allows us to find the explicit Shape Dynamics Hamiltonian through a phase space reduction $(\phi,\pi_\phi)\to(\phi_0,0)$. To find this map we can use $\pi_\phi\equiv 0$, so the $Q$ constraints become
\begin{equation}
 Q(\rho)=\int d^2 x \rho(x)\left(\pi(x)-\langle \pi \rangle\sqrt{|g|(x)}\right),
\end{equation}
which implies that $\pi(x)$ is a covariant constant. Using this and $\pi_\phi=0$, we find that the diffeomorphism constraint implies that
\begin{equation}
  PY_{ab}(x)=0,
\end{equation}
which implies that the scalar constraint is independent of $Y_a(x)$. Using these simplifications in the scalar constraints, we find
\begin{equation}\label{equ:SimpleScalarConstraint}
    S(N)=\int d^2 x \sqrt{|\bar g|}N \left(e^{-2(\hat \phi+\lambda)}\frac{\bar g_{ac}\bar g_{bd}p^{ab}p^{cd}}{|\bar g|}-\frac 1 2 e^{2(\hat \phi+\lambda)}(\langle \pi\rangle^2-4\Lambda)+2\bar \Delta (\hat \phi+\lambda)\right).
\end{equation}
Using (\ref{equ:standardMetricTorus}) and (\ref{equ:standardMomentumTorus}) we find that
\begin{equation}
\frac{\bar g_{ac}\bar g_{bd}p^{ab}p^{cd}}{|\bar g|} = \frac{\tau_2^2}{2}(p_1^2+p_2^2)
\end{equation}
is a spatial constant in the chosen chart. 

We see that the constraints $S(N)$ would be solved if we were able to choose $e^{4(\hat\phi+\lambda)}=\frac{2\bar g_{ac}\bar g_{bd}p^{ab}p^{cd}}{|\bar g|\left(\langle \pi\rangle^2-4\Lambda\right)}$. However, this is in general obstructed by the volume-preservation condition $\int d^2 x \sqrt{|\bar g|}e^{2(\hat \phi+\lambda)}=V$. This means that the constraints generating the refoliations are not completely gauge fixed by the condition $\pi_\phi(x)=0$. Indeed it turns out that among the infinitely many constraints $S(N)$ one remains first class, which after phase space reduction becomes our Shape Dynamics Hamiltonian $H_{SD}$. More concretely, there exists a lapse $N_0$ such that $S(N_0)$ Poisson commutes with $\pi_\phi$, i.e. it satisfies the lapse fixing equation
\begin{equation}\label{equ:LapseFixing}
 \begin{array}{rcl}
   \{S(N_0),\pi_\phi(x)\}&=&F_{N_0}(x)-e^{2(\hat\phi+\lambda)}\sqrt{|\bar g|}\langle F_{N_0}\rangle=0\\
   \textrm{where }F_{N}&=&N\left(-2e^{-2(\hat \phi+\lambda)}\bar g_{ac}\bar g_{bd}p^{ab}p^{cd}-|\bar g|e^{2(\hat \phi+\lambda)}(\langle \pi\rangle^2-4\Lambda)\right)+\sqrt{|\bar g|}\bar \Delta N.
 \end{array} 
\end{equation}
If one imposes on $N_0$ a normalization condition $\int d^2x \sqrt{|\bar g|}e^{2(\hat \phi+\lambda)} N_0=V$, then (\ref{equ:LapseFixing}) has a unique solution. We want to project out this first class part $S(N_0)$ from the full set of constraints $S(x)$ to end up with a purely second class set of constraints $\widetilde{S}(x)$ that we can solve. We can perform the projection in different ways, but a particularly convenient way of doing this is by defining
\begin{equation}
  \widetilde{S}(x):=S(x)-\frac{S(N_0)}{V}\sqrt{|\bar g|(x)}e^{2(\hat \phi(x)+\lambda(x))},
\end{equation}
which automatically satisfies $\widetilde{S}(N_0)=0$. Identifying $H_{SD}=S(N_0)$, we arrive at the modified Lichnerowicz-York equations
\begin{equation}\label{equ:LYtorus}
 \begin{array}{rcl}
    0&=&\widetilde{S}(x)=\sqrt{|\bar g|}\left(e^{-2(\hat \phi+\lambda)}\frac{\bar g_{ac}\bar g_{bd}p^{ab}p^{cd}}{|\bar g|}-\frac 1 2 e^{2(\hat \phi+\lambda)}(\langle \pi\rangle^2-4\Lambda+2\frac{H_{SD}}{V})-2\bar \Delta (\hat \phi+\lambda)\right),\\
    V&=&\int d^2 x \sqrt{|\bar g|}e^{2(\hat \phi+\lambda)},
  \end{array} 
\end{equation}
which we need to solve for $\hat{\phi}$ and $H_{SD}$. A solution is found by taking $\hat{\phi}+\lambda$ to be spatially constant. More precisely, from the second equation it follows that
\begin{equation}
\hat{\phi}=-\lambda + \frac 1 2 \ln V.
\end{equation}
Now $H_{SD}$ can be easily determined from the first equation in (\ref{equ:LYtorus}),
\begin{equation}\label{equ:SDHamiltonianTorus}
 H_{SD}=\frac 1 V \frac{\bar g_{ac}\bar g_{bd} p^{ab}p^{cd}}{|\bar g|} -\frac V 2  \left(\langle \pi\rangle^2-4\Lambda\right)=\frac{\tau_2^2}{2 V}(p_1^2+p_2^2)-\frac V 2  \left(\langle \pi\rangle^2-4\Lambda\right).
\end{equation}
Notice that to find $H_{SD}=S(N_0)$ we did not have to solve the lapse fixing equation explicitly. In this case we can solve (\ref{equ:LapseFixing}) straightforwardly using the fact that $\hat\phi +\lambda$ is constant and the result is simply $N_0=1$. In general however the lapse fixing equation is quite complicated and we are lucky that we don't actually have to solve it to derive the Shape Dynamics Hamiltonian (as we will again see in section \ref{sec:HigherGenus}). As a matter of fact, as a constraint $\widetilde{S}=0$ is completely equivalent to 
\begin{equation}\label{equ:TildeSAlternative}
S(x)-\langle S\rangle \sqrt{|\bar g|(x)}e^{2(\hat \phi(x)+\lambda(x))}=0,
\end{equation}
which does not refer to a lapse at all.

The Shape Dynamics Hamiltonian $H_{SD}$ (\ref{equ:SDHamiltonianTorus}) is exactly the reduced phase space Hamiltonian constraint. The more familiar Hamiltonian $H_{\textrm{York}}$ generating evolution in York time $\langle \pi \rangle$ (see e.g. \cite{Carlip:book} section 3.3) is obtained by noting that the variable canonically conjugate to $\langle \pi \rangle$ is $V$ and therefore by solving $H_{SD}=0$,
\begin{equation}
H_{\textrm{York}} = V = \tau_2\frac{\sqrt{p_1^2+p_2^2}}{\sqrt{\langle\pi\rangle^2-4\Lambda}}.
\end{equation} 

We can now perform explicitly the phase space reduction of the linking theory and describe Shape Dynamics on the ADM phase space through its total Hamiltonian and first class constraints
\begin{equation}\label{equ:SDontorus}
 \begin{array}{rcl}
   H&=&\mathcal N H_{SD}+H(\xi)+C(\rho)\\
   H_{SD}&=&\frac{\tau_2^2}{2V}(p_1^2+p_2^2) -\frac V 2  \left(\langle \pi\rangle^2-4\Lambda\right)\\
   H(\xi)&=&\int d^2 x \pi^{ab}\mathcal L_\xi g_{ab}\\
   C(\rho)&=&\int d^2 x \rho\left(\pi-\langle \pi\rangle\sqrt{|g|}\right).
 \end{array}
\end{equation}
The gauge symmetries are indeed spatial diffeomorphisms, conformal transformations that preserve the total volume and global time reparametrizations. Despite the different set of symmetries, the equivalence with standard General Relativity is obvious: The Shape Dynamics Hamiltonian coincides on the reduced phase phase with the CMC Hamiltonian, while the constraints $C$ provide the CMC gauge-fixing conditions.

Although we know the Shape Dynamics Hamiltonian explicitly on the torus, it is instructive to observe that the Shape Dynamics Hamiltonian constraint $H$ can be expanded in powers of the inverse volume, because it shows two properties that we can investigate in more complicated models. This expansion is a systematic approximation to Shape Dynamics that is a good approximation in an asymptotic large volume regime, i.e. where $V\to\infty$ while keeping the other degrees of freedom finite. In this regime we find two important features of Shape Dynamics:
\begin{enumerate}
 \item  {\bf Asymptotic Locality:} The leading order of the Hamiltonian, which becomes exact in the limit $V\to \infty$, is $\langle\pi\rangle^2-4\Lambda+\mathcal O(V^{-2})\approx 0$. As a constraint, this is equivalent to 
   \begin{equation}\label{equ:asymptoticHSD}
     V\left(\langle\pi\rangle-2\sqrt{\Lambda}\right)=\int d^2x\left( \pi(x)-2\sqrt{\Lambda}\sqrt{|g|}(x) \right)\approx 0\,\,\,\textrm{ for }V\to\infty
   \end{equation} 
   which is diffeomorphism invariant as the integral over a {\emph{local}} density and by inspection invariant under conformal transformations that preserve the total volume. 
 \item {\bf Full Conformal Invariance:} Since the Shape Dynamics Hamiltonian constraint is asymptotically equivalent to $\langle\pi\rangle-const.\approx 0$, we can add it to the conformal constraints $C$ to obtain in the large volume limit $C(x)+H_{SD}=\pi(x)-const.\approx 0$, which generates full conformal transformations, i.e., including those that change the total spatial volume. Notice that this requires us to interpret the Shape Dynamics Hamiltonian as a constraint, rather than a generator of physical dynamics.
\end{enumerate}
Let us have a quick look at the {\bf 2-sphere}: The linking theory and phase space reduction can be performed following the same steps as on the torus with two small modifications. 1) there are no Teichm\"uller parameters on the sphere, so there is only one canonical pair of physical degrees of freedom ($V$ and $\langle \pi\rangle$); 2) the total spatial curvature does not vanish, but is $8\pi$. Shape Dynamics on the sphere thus takes the form of equation (\ref{equ:SDontorus}), except for the Hamiltonian, which is $H_{SD}=-\frac V 2 (\langle \pi\rangle^2-4\Lambda)-8\pi$.

\section{Higher Genus Surfaces}\label{sec:HigherGenus}

On a higher genus surface, we can still use the decomposition analogous to equations (\ref{equ:metricDecomposition}) and (\ref{equ:momentumDecomposition}), but the explicit construction of the Shape Dynamics Hamiltonian constraint on the torus rested on the explicit solvability of the modified Lichnerowicz--York equation (\ref{equ:LYtorus}). The Lichnerowicz--York equation on a higher genus surface (or in higher dimensions) is not explicitly solvable. We thus restrict our construction of Shape Dynamics to an approximation scheme and consider an expansion that becomes exact in the large volume limit.\footnote{See \cite{Banks:1984np} for an analogous expansion.} Again, we follow \cite{Carlip:2004ba,Carlip:book} when possible.

\subsection{Preparations}

In genus $g\geq 2$ we can perform a decomposition analogous to (\ref{equ:metricDecomposition}) and (\ref{equ:momentumDecomposition}),
\begin{equation}\label{equ:HigherGenusDecomposition}
  \begin{array}{rcl}
    g_{ab}(x)&=&e^{2\lambda(x)}(f^* \bar g)_{ab}(x),\\
    \pi^{ab}(x)&=&e^{-2\lambda}\left(p^{ab}(x)+\frac 1 2 \bar g^{ab}(x)\pi(x)+\sqrt{|\bar g|}(x)\bar g^{ab}(x)\bar g^{cd}(x)PY_{cd}(x)\right),
  \end{array}
\end{equation}
where now we take the reference metric $\bar{g}$ to be of unit volume $\int d^2x\sqrt{|\bar g|}=1$ and constant scalar curvature $\bar{R}$. According to the Gauss--Bonnet theorem, $\bar R$ is given by
\begin{equation}
\bar{R} = - 8\pi (g-1).
\end{equation}
Modulo diffeomorphisms the space of such metrics corresponds to the genus $g$ Teichm\"uller space, which has dimension $6g-6$. Unfortunately no simple explicit parametrization for $\bar{g}$ is known, so we will keep the parametrization implicit. 

We can again write the linking theory using the decomposition (\ref{equ:HigherGenusDecomposition}). The only difference compared to the constraints (\ref{equ:torusConstraints}) for the torus is the subtraction from $S(N)$ of a spatial curvature term.

When we impose the gauge fixing $\pi_\phi=0$, we obtain the analogue of (\ref{equ:SimpleScalarConstraint}):
\begin{equation}
    S(N)=\int d^2 x \sqrt{|\bar g|}N \left(e^{-2(\hat \phi+\lambda)}\frac{\bar g_{ac}\bar g_{bd}p^{ab}p^{cd}}{|\bar g|}-\frac 1 2 e^{2(\hat \phi+\lambda)}(\langle \pi\rangle^2-4\Lambda)+2\bar \Delta (\hat \phi+\lambda)-\bar R\right).
\end{equation}
Reusing our discussion for the torus, we construct the second class part $\widetilde S$ of $S$ according to (\ref{equ:TildeSAlternative}),
\begin{equation}\label{equ:TildeSHigherGenus}
\widetilde{S}(x)=S(x)-\langle S\rangle \sqrt{|\bar g|(x)}e^{2(\hat \phi(x)+\lambda(x))}.
\end{equation}
Identifying the remaining first class constraint $\langle S\rangle$ with $H_{SD}/V$, we obtain the modified Lichnerowicz--York equations for genus $g \geq 2$,
\begin{equation}\label{equ:LYhighergenus}
 \begin{array}{rcl}
    0&=&\widetilde{S}(x)=\sqrt{|\bar g|}\left(e^{-2(\hat \phi+\lambda)}\frac{\bar g_{ac}\bar g_{bd}p^{ab}p^{cd}}{|\bar g|}-\frac 1 2 e^{2(\hat \phi+\lambda)}(\langle \pi\rangle^2-4\Lambda+2\frac{H_{SD}}{V})+2\bar \Delta (\hat \phi+\lambda)- \bar R\right),\\
    V&=&\int d^2 x \sqrt{|\bar g|}e^{2(\hat \phi+\lambda)}.
  \end{array} 
\end{equation}
To simplify the notation let us define $\mu=\hat\phi+\lambda-\frac{1}{2}\ln V$ and $p^2 := \frac{\bar g_{ac}\bar g_{bd}p^{ab}p^{cd}}{|\bar{g}|}$. Then (\ref{equ:LYhighergenus}) can be written
\begin{equation}\label{equ:LYsimplified}
\frac{1}{V}p^2e^{-2\mu} - \frac{1}{2} V \left(\langle \pi\rangle^2-4\Lambda+\frac{2 H_{SD}}{V}\right) e^{2\mu}+ 2 \bar{\Delta} \mu - \bar R= 0 \quad \textrm{and}\quad \langle e^{2\mu}\rangle_{\bar{g}} = 1.
\end{equation}
In the following we will drop the subscript $\bar{g}$ and keep in mind that averages $\langle\cdot\rangle$ are taken with respect to $\bar{g}$ (except for $\langle\pi\rangle$).

Equation (\ref{equ:LYsimplified}) is nearly identical to the standard Lichnerowicz--York equation in 2+1 dimensions. The only difference is that we have a restriction on $\mu$ and to compensate this we have an additional constant $H_{SD}$ to solve for. The existence of a unique solution for $\mu$ and $H_{SD}$ (as a function of $\bar{g}_{ab}$, $p^{ab}$, $V$ and $\langle \pi \rangle$) is a direct consequence of the existence and uniqueness properties of the usual Lichnerowicz--York equation.

The key simplification that allowed us to explicitly construct Shape Dynamics on the torus is that there one can choose the constant curvature metric $\bar{g}_{ab}$ such that $\bar g_{ac}\bar g_{bd}p^{ab}p^{cd}/|\bar{g}|$ is spatially constant (as is apparent from  (\ref{equ:metricDecomposition})). For genus $2$ and higher the LY equation is much harder to solve. However, we can already deduce some properties of $H_{SD}$ by integrating expression (\ref{equ:LYsimplified}),
\begin{equation}\label{equ:GeneralFormHgf}
H_{SD} = -\frac{V}{2}(\langle \pi\rangle^2-4\Lambda) - \bar{R} + \frac{1}{V} \langle p^2 e^{-2\mu}\rangle.
\end{equation}
We have chosen our second class constraints (\ref{equ:TildeSHigherGenus}) in such a way that the solution $\mu$ will not depend on $\langle\pi\rangle$ (or $\Lambda$), and therefore the same holds for the last term in (\ref{equ:GeneralFormHgf}). Hence, our choice is special in that it produces a Hamiltonian that is quadratic in the momentum $\langle \pi\rangle$ conjugate to $V$.

\subsection{Large Volume Asymptotic Expansion}\label{sec:LargeVExpansion}

Although we can not solve (\ref{equ:LYsimplified}) explicitly, our modified LY equation does allow for an interesting perturbative expansion. Indeed notice that the volume $V$ appears explicitly in (\ref{equ:LYsimplified}) and should be treated as a parameter when solving the equation. Therefore we can try to find solutions $\mu$ and $H_{SD}$ expanded in powers of $1/V$ and construct the Shape Dynamics in the infinite volume limit.\footnote{Notice that this can not easily be done in the reduced phase space approach, in which also the volume itself has to be solved for in terms of $\bar{g}$ and the momenta.} To do this we make the ansatz
\begin{equation}
e^{2\mu} = \sum_{k=0}^{\infty} \Omega_k V^{-k} \,\,\,\textrm{ and }\,\,\, H_{SD} =  \sum_{k=-1}^{\infty} H_k V^{-k}.
\end{equation} 
Indeed, from carefully looking at equation (\ref{equ:LYsimplified}) it follows that higher powers of $V$ can not occur. From the normalization $\langle e^{2\mu}\rangle=1$ we get the restrictions $\langle\Omega_0\rangle = 1$ and $\langle\Omega_k\rangle = 0$ for $k > 0$.  

The leading order of equation (\ref{equ:LYsimplified}) is proportional to $V$ and fixes $H_{-1} = -\frac{1}{2}(\langle \pi\rangle^2-4\Lambda)$. At order $V^0$ the equation then reads
\begin{equation}
-\Omega_0 H_0-\bar{R}+2\bar{\Delta}\ln(\Omega_0) = 0,
\end{equation}
which is clearly solved by $H_0 = -\bar{R}$ and $\Omega_0=1$. The LY equation now becomes
\begin{equation}\label{equ:LYExpandedInV}
\frac{1}{V}\frac{1}{1+\frac{\Omega_1}{V}+\cdots} p^2 - \left(1+\frac{\Omega_1}{V}+\cdots\right)\left(-\bar{R} + \frac{H_1}{V}+\frac{H_2}{V^2}+\cdots\right) - \bar{R} + \bar{\Delta}\ln\left(1+\frac{\Omega_1}{V}+\cdots\right)=0.
\end{equation}
If we define the polynomials $A_k$ and $B_k$ in $\Omega_1$ through $\Omega_k$ by\footnote{The first few polynomials are $A_0=1$, $A_1 = - \Omega_1$, $A_2 = -\Omega_2+\Omega_1^2$, $A_3 = -\Omega_3+2\Omega_1\Omega_2-\Omega_1^3$ and $B_0=0$,$B_1=-\Omega_1^2/2$, $B_2=\Omega_1^3/3 - \Omega_1\Omega_2$.}
\begin{equation}
\frac{1}{1+\sum_k\frac{\Omega_k}{V^k}} = \sum_{k} \frac{A_k[\Omega]}{V^k}  \,\,\,\textrm{ and }\,\,\, \ln(1+\sum_k\frac{\Omega_k}{V^k})-\sum_k\frac{\Omega_k}{V^k} = \frac{1}{V}\sum_{k} \frac{B_k[\Omega]}{V^k}
\end{equation}
then from equation (\ref{equ:GeneralFormHgf}) or from integrating (\ref{equ:LYExpandedInV}) it follows that for $k \geq 1$
\begin{equation}\label{equ:HkfromOmega}
H_k = \left\langle p^2A_{k-1}[\Omega] \right\rangle.
\end{equation}
The order $V^{-k}$ equation in (\ref{equ:LYExpandedInV}) then allows us to solve $\Omega_k$ explicitly in terms of $\Omega_0$ through $\Omega_{k-1}$,
\begin{equation}\label{equ:RecurrenceForOmega}
\Omega_k = (\bar{\Delta}+\bar{R})^{-1}\left(- A_{k-1}[\Omega] p^2 - \bar{\Delta}B_{k-1}[\Omega]+ \sum_{l=1}^{k}H_l\Omega_{k-l}\right),
\end{equation}
where the operator $\bar{\Delta}+\bar{R}$ is negative definite and therefore has a well-defined inverse.

We have therefore obtained a general algorithm to solve the modified LY equation order by order through the recurrence relation (\ref{equ:RecurrenceForOmega}) together with (\ref{equ:HkfromOmega}). We have calculated the first few $H_k$ explicitly leading to a Hamiltonian
\begin{equation}\label{equ:SDHamiltoninInVExpansion}
\begin{array}{rcl}
H_{SD} &=& -\frac{V}{2}(\langle \pi\rangle^2-4\Lambda) - \bar{R} + \frac{1}{V} \langle p^2 \rangle + \frac{1}{V^2} \left\langle ( p^2 - \langle p^2\rangle )(\bar{\Delta} + \bar{R})^{-1}( p^2 - \langle p^2\rangle )\right\rangle \\
&& +\frac{1}{2 V^3}\left[ k \left\langle ((\bar{\Delta}+\bar{R})^{-1}(p^2-\langle p^2\rangle))^3\right\rangle + 3 \left\langle (p^2+\langle p^2\rangle)((\bar{\Delta}+\bar{R})^{-1}(p^2-\langle p^2\rangle))^2\right\rangle\right]+\cdots.
\end{array}
\end{equation}  
In general $H_k$ will be a function homogeneous in $p^2$ of order $k$.\footnote{As the $H_k$ are actually functions on the cotangent bundle to Teichm\"uller space, one might ask how natural they are from the perspective of Teichm\"uller spaces. As a partial answer we notice that $H_1=\langle p^2\rangle$ is related to the canonical Weil--Petersson metric \cite{Wolpert:2008wp,Takhtajan:2003hm}, while $H_2$ is closely related to its curvature \cite{Wolpert:2010wp}.} 

The expansion has features similar to a tree level (Feynman diagram) expansion with propagator $(\bar{\Delta}+\bar{R})^{-1}$ and source term $p^2/V$. To make this connection more explicit, let us view the modified LY equation (\ref{equ:LYsimplified}) as the Euler--Lagrange equation of some action $S_{LY}$. Such an action can be easily constructed,
\begin{equation}\label{equ:lyaction}
S_{LY}[\mu,H] = \int d^2x \sqrt{\bar{g}} \left( \mu\bar{\Delta}\mu - \bar{R} \mu - (H-\frac{\bar{R}}{2}) \left(e^{2\mu}-1\right) - \frac{1}{2} \frac{e^{-2\mu}}{V} p^2 \right),
\end{equation}
where $\mu$ is now viewed as an unrestricted function since the Lagrange multiplier $H = \frac{1}{2}(H_{SD}+\frac{V}{2} (\langle \pi \rangle^2 - 4 \Lambda)+\bar{R})$ enforces the constraint $\langle e^{2\mu}\rangle=1$ on variation. We can rewrite (\ref{equ:lyaction}) by singling out the quadratic part  in $\mu$ and $H$, 
\begin{equation}
S_{LY}[\mu,H] = \int d^2x \sqrt{\bar{g}} \left( \mu(\bar{\Delta}+\bar{R})\mu- 2 \mu H + \bar{R} (\frac{2}{3} \mu^3 + \frac{1}{3} \mu^4+\ldots) - H(2\mu^2+\ldots) - \frac{1}{2} \frac{e^{-2\mu}}{V} p^2 \right).
\end{equation}
The Feynman rules can be read off and we can find $H$ (and therefore $H_{SD}$) by computing its tree-level one-point function. Notice that the action (\ref{equ:lyaction})  which we use to derive the Hamiltonian is similar to that of two dimensional Liouville gravity \cite{Ginsparg:1993is}. More precisely, it is of the form of a Liouville action plus a perturbation given by a source term proportional to $p^2/V$. 

Two remarks are in order:
\begin{enumerate}
  \item We observe from equation (\ref{equ:GeneralFormHgf}) that in the limit $V\to\infty$ the Shape Dynamics Hamiltonian again approaches a form that is equivalent to (\ref{equ:asymptoticHSD}). The Hamiltonian is thus asymptotically local and provides the volume-changing generator of conformal transformations, so full conformal invariance is asymptotically attained.
  \item The first three terms in the large volume expansion (\ref{equ:SDHamiltoninInVExpansion}) sum up to an expression equivalent to the temporal gauge Hamiltonian $S(N\equiv 1)$, which is {\emph{local}}.
\end{enumerate}

\section{Dirac Quantization in Metric Variables}\label{sec:Quantize}

To expose the difference between Shape Dynamics and General Relativity, we consider the Dirac Quantization of pure gravity on the torus in 2+1 dimensions in metric variables, usually referred to as Wheeler--DeWitt quantization. For the sake of completeness we first follow \cite{Carlip:book} and revisit the problems associated with the nonlocality arising from the solution of the diffeomorphism constraint in the Wheeler--DeWitt approach. Subsequently we show how these problems are solved by trading local Hamiltonian constraints for local conformal constraints, which allows us to perform a Dirac Quantization program for Shape Dynamics.

\subsection{Dirac Quantization of General Relativity on the 2+1 Torus}

Contrary to first order variables, one can not readily quantize General Relativity in metric variables even in 2+1 dimensions on the torus and sphere. The reason is well explained in \cite{Carlip:book}, which we follow here. Using the standard decomposition of the metric and momenta, we can solve the diffeomorphism- constraints for the transverse part of the momenta
\begin{equation}
  \bar Y_i=-\frac 1 2 \left(\bar \Delta+\frac k 2\right)^{-1}\left(e^{2\lambda}\bar \nabla_i\left(e^{-2\lambda}\pi\right)\right),
\end{equation}
where $k=0$ for the torus. To perform a Wheeler--DeWitt quantization reference \cite{Carlip:book} chooses a polarization for which the configuration operators are given by functionals of the spatial metric and formally considers a Schr\"odinger representation on wave functions $\psi[f,\lambda;\tau)$, which reduces to $\psi[\lambda;\tau)$. Assuming that the inner product is constructed from a divergence-free measure, we can quantize the momenta by replacing $\pi\to -\frac i 2 \frac{\delta}{\delta\lambda}$ and $p^{ab}\to -i\frac{\partial}{\partial\hat g_{ab}}$. The diffeomorphism constraint on the torus still acts nontrivially on the conformal factor, and its solution can be quantized as
\begin{equation}
 \bar Y_i[\hat \pi]=\frac i 4 \bar \Delta^{-1}\left(e^{2\lambda}\bar \nabla_i\left(e^{-2\lambda}\frac{\delta}{\delta \lambda}\right)\right).
\end{equation}
This expression is plugged into the Hamiltonian and leads to nonlocal terms in the Wheeler--DeWitt equation that are not practically manageable and lead to notorious difficulties in the Wheeler--DeWitt approach \cite{Carlip:1993ak}. To make a connection with reduced phase space quantization reference \cite{Carlip:book} assumes a solution $\psi_o[\lambda;\tau)$ to the Wheeler--DeWitt equation and restricts it to constant York time $T$ through
\begin{equation}
 \hat \psi_o(T,\tau):=\int \mathcal D\lambda e^{i\,T\,\int d^2 x e^{2\lambda}}\psi_o[\lambda;\tau)
\end{equation}
and inserts this into the Wheeler--DeWitt equation (with vanishing cosmological constant). In terms of the Teichm\"uller Laplacian $\Delta_o:=-\tau_2^2\left(\partial^2_{\tau_1}+\partial^2_{\tau_2}\right)$ this yields
\begin{equation}
 \left(\left(T\frac \partial {\partial T}\right)^2+\Delta_o\right)\hat\psi_o(T,\tau)=\int D\lambda e^{iT\int d^2 x e^{2\lambda}}\left(T^2\left(e^{4\lambda}-V^2\right)+4e^{2\lambda}\bar\Delta \lambda\right)\psi_o[\lambda;\tau),
\end{equation}
where the RHS vanishes if $\psi_o$ has support only on spatially constant conformal factors, while the LHS is equivalent to a reduced phase space quantization. 

\subsection{Dirac Quantization of Shape Dynamics on the 2+1 Torus}

We now follow essentially the same strategy as in the previous subsection but for Shape Dynamics. We choose a polarization where functionals of the metric are configuration variables and formally consider a Schr\"odinger representation on functionals $\psi[\lambda,f;\tau)$, such that functionals of the metric are represented by multiplication operators and also re-use the representation of the momentum operators given in the previous section. We start with the local conformal constraint $\pi(x)-\frac{\sqrt{|\bar g|}e^{2\lambda(x)}}{V}\int d^2 y \pi(y)=0$, which can be readily quantized as
\begin{equation}
  -\frac i 2\left(\frac{\delta}{\delta \lambda(x)}-\frac{\sqrt{|\bar g|}e^{2\lambda(x)}}{V}\int d^2y\frac{\delta}{\delta \lambda(y)}\right)\psi[\lambda,f;\tau)=0,
\end{equation}
where we work in a chart where the components of $\bar g_{ab}$ are constant. The solution to this constraint is that $\psi$ depends only on the homogeneous mode of $\lambda$. We can thus write the general solution to the local conformal constraints as a wave function of $\psi[f;V,\tau)$, where $V$ denotes the spatial volume. We now turn to the spatial diffeomorphism constraint. Exponentiating the spatial diffeomorphism constraint to finite diffeomorphisms implies that for each small diffeomorphism $f_o$ there is a unitary operator acting as the pull-back under a diffeomorphism:
\begin{equation}
 U_{f_o}\psi: [\lambda,f;\tau)\mapsto \psi[f_o^*\lambda,f_o\circ f;\tau),
\end{equation}
where we assume that the measure is diffeomorphism invariant, so $U_{f_o}$ is indeed unitary. The pull-back action $f_o^* \lambda$ on the conformal factor is the source of the nonlocal terms that we encountered in the action of the diffeomorphisms in the previous subsection. This action is however trivial on the space of solutions to the local conformal constraint, since $f_o^*V=V$. We can thus easily impose the diffeomorphism constraint
\begin{equation}
 U_{f_o} \psi = \psi
\end{equation}
for all diffeomorphisms $f_o$, which implies for solutions to the local conformal constraint that $\psi[f;V,\tau)$ is independent of $f$. We thus find that the solution space to the local constraints of Shape Dynamics consists of Schr\"odinger wave functions $\psi(V,\tau)$. We have not specified the kinematic Hilbert-space that we used, since we returned to reduced phase space Hilbert space by solving the linear constraints at the quantum level.\footnote{We expect that a rigorous construction of the kinematic Hilbert space is possible as follows: First we specify a measure $d\mu$ on $\mathbb R^3$ and consider the Hilbert space $\mathcal H_o=L^2(\mathbb R^3,d\mu)/a.e.$ of square integrable functions modulo the set of functions that have support only on sets of measure zero. Next, we construct a Hilbert space $\mathcal H_d$ as the closure of $\oplus_{f\in Diff.} \mathcal H_f$, where $Diff.$ denotes the group of diffeomorphisms that are not isometries of reference metrics and where all $\mathcal H_f$ have one complex dimension. Then we construct a Hilbert space $\mathcal H_c$ as the closure of $\oplus_{\lambda \in VPCT} \mathcal H_{\hat \lambda}$, where $\hat \lambda$ is a conformal factor that preserves the total volume of all standard metrics (this is possible since the volume element of the standard metrics does not depend on the point in Teichm\"uller space of the 2-torus) and where again each $\mathcal H_{\hat \lambda}$ has one complex dimension. We define our representation now on the tensor product $\mathcal H=\mathcal H_o \otimes \mathcal H_d \otimes \mathcal H_c$. Informally speaking the Hilbert spaces $\mathcal H_d$ and $\mathcal H_c$ consist of the closed span of normalized eigenfunctions of the diffeomorphism resp. conformal degrees of freedom of the metric. The reason for this construction is that we want to solve the diffeomorphism and local conformal constraints by group averaging, which collapses the space $\mathcal H_d\otimes \mathcal H_c$ to $\mathbb C$. Solving these constraints thus leads to the Hilbert space $\mathcal H_o$.}

We now consider the Shape Dynamics Hamiltonian $H_{SD}=\tau_2^2\left(p_1^2+p_2^2\right)-V\left(\langle\pi\rangle^2-4\Lambda\right)$, which can be quantized on the factor $\mathcal H_o$ that remains after solving the linear constraints by replacing $p_i \to -i\frac{\partial}{\partial \tau^i}$ and $\langle \pi\rangle \to -i \frac{\partial}{\partial V}$. This leads to the quantum Shape Dynamics Hamiltonian
\begin{equation}
 H=-\tau_2^2\left(\partial^2_{\tau_1}+\partial^2_{\tau_2}\right)+V^2\left(\partial^2_V+4\Lambda\right).
\end{equation}
This is the covariant reduced phase space Hamiltonian \cite{Martinec:1984fs}. We thus confirmed the expectation of the last subsection that Dirac Quantization of Shape Dynamics should be equivalent to reduced phase space quantization.

\subsubsection*{Excursion: Large Volume Behavior of the Quantum Theory}

We saw in the classical theory that full conformal invariance is attained in the large volume regime. Let us now investigate this question in the quantum theory. This is particularly interesting in light of a large-CMC-volume/conformal field theory correspondence. For a first investigation we neglect the issue of modular invariance and simply investigate the asymptotic volume dependence of solutions $\psi(V,\tau_1,\tau_2)$ to the Shape Dynamics Wheeler--DeWitt equation 
\begin{equation}
  \left(-\tau_2^2\left(\partial^2_{\tau_1}+\partial^2_{\tau_2}\right)+V^2\left(\partial^2_V+4\Lambda\right)\right)\psi(V,\tau_1,\tau_2)=0.
\end{equation}
We can use a separation ansatz $\psi(V,\tau_1,\tau_2)=v(V)m(\tau_1,\tau_2)$ and introduce separation constants $\alpha$. This implies $\tau_2^2\left(\partial^2_{\tau_1}+\partial^2_{\tau_2}\right) m(\tau_1,\tau_2)=\alpha m(\tau_1,\tau_2)$ and
\begin{equation}
  v^{\prime\prime}(V)+ \left(4 \Lambda -\frac{\alpha}{V^2}\right)v(V)=0,
\end{equation}
which is solved by $v_\pm(V)=\sqrt{V\pi\sqrt{\Lambda}}e^{\pm i\frac{\pi}{2}\left(\frac 1 2+\sqrt{\alpha+1/4}\right)}\left(J_{\sqrt{\alpha+1/4}}(2\sqrt{\Lambda}V)\pm i Y_{\sqrt{\alpha+1/4}}(2\sqrt{\Lambda}V)\right)$ in terms of Bessel functions $J_\nu,Y_\nu$. We are interested in the limit $V\to\infty$, where the term $\frac{\alpha}{V^2}$ vanishes and the two linearly independent asymptotic solutions are
\begin{equation}
 v_\pm(V)=e^{\pm i 2\sqrt{\Lambda} V} \left(1+\mathcal O\left(\frac 1 V\right)\right) \,\,\textrm{ for }\,V\to \infty,
\end{equation}
which is at leading order {\emph{independent}} of the separation constant $\alpha$ and hence asymptotically true for all solutions to the Shape Dynamics Wheeler--DeWitt equation. The general asymptotic scaling under global conformal transformations $C=-i V\partial_V$ is thus
\begin{equation}\label{equ:globalScaling}
  C v_\pm(V)= \pm 2 \sqrt{\Lambda} V v_\pm(V) \,\,\textrm{ for } V\to \infty.
\end{equation}
The difference between the Shape Dynamics wave function and the reduced phase space quantization wave function is that the Shape Dynamics wave function is a function of the full metric $\psi[g]$ that is constrained to be independent of the local degrees of freedom (by diffeomorphism- and local conformal invariance) and is thus completely specified by its dependence on $V,\tau_1,\tau_2$, while the reduced phase space quantization wave function is a function  $\psi(V,\tau_1,\tau_2)$ on reduced phase space only. We thus have the asymptotic scaling of a generic Shape Dynamics wave function under global conformal transformations
\begin{equation}
 C \psi_\pm[g]= \pm 2 \sqrt{\Lambda} V \psi_\pm[g] \,\,\textrm{ for } V\to \infty,
\end{equation}
which can be combined with manifest invariance of $\psi[g]$ under volume preserving conformal transformations $ C_{VPCT}(x) \psi_\pm[g]=0 $ to yield the asymptotic scaling of $\psi[g]$ under arbitrary conformal transformations
\begin{equation}
  C(x) \psi_\pm[g]=\pm 2 \sqrt{\Lambda} \sqrt{|g|(x)} \psi_\pm[g] \,\,\textrm{ for } V\to \infty.
\end{equation} 
Notice the explicit deviations from this scaling at order $1/V$, which can be derived by straightforward series expansion of $v_\pm(V)$. Compare this with \cite{Freidel:2008sh} where a similar result is discussed in the context of the AdS/CFT-correspondence and \cite{Gomes:2011dc} for a large CMC-volume/CFT discussion.

\section{Possible Consequences for Shape Dynamics in Higher Dimensions}\label{sec:HigherDimensions}

Pure gravity in 2+1 dimensions does not possess local degrees of freedom, which is the technical reason for many explicit constructions that can not be performed in higher dimensions. However, having an example that can be worked out explicitly can give guidance for the treatment of higher dimensions. It is the purpose of this section to explore what consequences can be drawn from Shape Dynamics in 2+1 dimensions, in particular from the torus universe. A natural generalization of the simplifications occurring on the 2+1 torus universe to 3+1 dimensions is the strong gravity or BKL-limit, where spatial derivatives in the ADM Hamilton constraints can be neglected, so
\begin{equation}
 S_{ADM}^{BKL}(x)=\frac{\pi^{ab}(x)G_{abcd}(x)\pi^{cd}(x)}{\sqrt{|g|(x)}}+2\Lambda\sqrt{|g|(x)}.
\end{equation}
In this case we can algebraically solve the analogue of the Lichnerowicz--York equation and $\langle e^{6\hat \phi}\rangle=1$ for the Shape Dynamics BKL Hamiltonian in 3+1 dimensions
\begin{equation}
 H_{SD}^{BKL}=V\left(\left\langle \sqrt{\sigma^{ab}g_{ac}g_{bd}\sigma^{cd}}\right\rangle_g^2-\frac 1 6 \langle \pi\rangle_g^2+2\Lambda\right),
\end{equation}
where $\sigma^{ab}=\pi^{ab}-\frac 1 3\pi g^{ab}$. We observe that the ingredients $V,\langle \sqrt{Tr(\sigma.\sigma)}\rangle$ and $\langle\pi\rangle$ are each invariant under volume preserving conformal transformations as well as invariant under spatial diffeomorphisms at the expense of nonlocality. The absence of spatial derivatives is what allows us to explicitly construct the BKL-Shape Dynamics Hamiltonian in higher dimensions, i.e. for a similar reason for which the Shape Dynamics Hamiltonian could be explicitly constructed for the spherical- and torus- universe in 2+1 dimensions.\footnote{The Lichnerowicz-York equation on the torus is algebraically solvable by applying the maximum principle, while the BKL Hamiltonian is algebraic to begin with.} 

In an effort to include spatial derivatives, we could decompose a 3-dimensional metric analogously into a conformal factor, diffeomorphism and reference metric to the decomposition in 2 dimensions. For this we observe that locally one can specify a diffeomorphism class of a metrics by giving three independent curvature invariants  e.g. $(\phi_1,\phi_2,\phi_3)=\left(R,R_{ab}R^{ab},\frac{|R|}{|g|}\right)$ and integration constants $\tau$. Using the Yamabe problem on a compact manifold without boundary, one can then impose $R(x)=\langle R\rangle$ as a gauge fixing condition for the conformal factor and thus locally find reference metrics $g_{ab}[\langle R\rangle,\phi_2,\phi_3,\tau;x)$. Despite this being a purely formal construction, since in contrast with the 2-torus and sphere the construction of reference metrics is not feasible as it requires the inversion of a complicated system of coupled partial differential equations, one finds that one can still not solve for the Shape Dynamics Hamiltonian in all phase space, because of the position dependence of $\frac{\sigma^{ab}\sigma_{ab}}{|g|}(x)$. Let us therefore consider the restricted phase space
\begin{equation}
 \Gamma_r=\left\{(g,\pi)\in \Gamma_{ADM}:\frac{\sigma^{ab}\sigma_{ab}}{|g|}(x)=\langle\frac{\sigma^{ab}\sigma_{ab}}{|g|}\rangle\,\,\text{and}\,\,R(x)=\langle R\rangle\right\}.
\end{equation}
Since we can use $R(x)=\langle R\rangle=:R_o$ as a gauge fixing for the conformal gauge symmetry on a compact manifold without boundary, we see that the condition $\frac{\sigma^{ab}\sigma_{ab}}{|g|}(x)=\langle\frac{\sigma^{ab}\sigma_{ab}}{|g|}\rangle$ constrains one local physical degree of freedom, so $\Gamma_r$ contains only 3/4 of the physical degrees of freedom of Shape Dynamics. The fact that the solution to the modified Lichnerowicz--York equation is homogeneous if all coefficient functions are homogeneous and that the volume preservation condition for the conformal factor implies that a homogeneous conformal factor is $\phi(x)=\frac{\ln(V)} 6$ lets us find
\begin{equation}
 \left.H_{SD}\right|_{\Gamma_r}=\int d^3 x\left(\frac{\pi^{ab}G_{abcd}\pi^{cd}}{\sqrt{|g|}}-\left(R-2\Lambda\right)\sqrt{|g|(x)}\right),
\end{equation}
i.e. the ADM-Hamiltonian with homogeneous lapse. Observing that on $\Gamma_r$ we have $\int d^3 x \frac{\pi^{ab}G_{abcd}\pi^{cd}}{\sqrt{|g|}}=V\left\langle \sqrt{\sigma^{ab}g_{ac}g_{bd}\sigma^{cd}}\right\rangle_g^2-\frac V 6 \langle \pi\rangle^2$ and $\int d^3x \sqrt{|g|}R=V R_o$, where $R_o$ denotes the Yamabe constant expressed as a nonlocal functional obtained by solving $R[e^{4\hat\lambda}g;x)=\langle R\rangle$ for $\lambda_o[g;x)$ and $R_o[g;x):=R[e^{4\hat\lambda[g;x)}g;x)$, we can immediately extend this to the conformal orbit $\Gamma_c$ of $\Gamma_r$ to yield
\begin{equation}
  \left.H_{SD}\right|_{\Gamma_c}=V\left(\left\langle \sqrt{\sigma^{ab}g_{ac}g_{bd}\sigma^{cd}}\right\rangle_g^2-\frac 1 6 \langle \pi\rangle^2-\langle R_o\rangle+2\Lambda\right),
\end{equation}
which differs from the Shape Dynamics Hamiltonian for 1/4 of the physical degrees of freedom. In summary, we see that one has to resort to approximation schemes to solve for the classical Shape Dynamics Hamiltonian that is equivalent to General Relativity, which is an obstruction to the direct construction of a quantum theory of Shape Dynamics. 

One may understand this as a hint to use the BKL-Hamiltonian as a starting point for Shape Dynamics and introduce spatial derivatives as perturbations in the classical theory. Similarly, one may hope to quantize the ultralocal theory and develop a perturbation scheme at the quantum level. The feasibility of this idea remains to be explored, but if possible it would lead to an expansion of the quantum theory that is exact precisely where quantum effects are expected to be most important.\footnote{The BKL-conjecture \cite{Belinsky:1970ew,Khalatnikov:1969eg,Belinsky:1982pk} roughly states that spatial derivatives are negligible when a generic singularity is approached, which is the regime where quantum effects of gravity are expected to be most important.}

A different direction could be to follow the original construction of Hamiltonians in Loop Quantum Gravity, where one first constructs a rigging map to solve the diffeomorphism constraint and subsequently constructs a quantization of the General Relativity Hamiltonian constraints on the solutions to the diffeomorphism constraint to avoid problems with anomalies. An analogous strategy for Shape Dynamics is to first construct a rigging map to solve the local conformal constraint and subsequently construct a quantization of the Shape Dynamics Hamiltonian on the solution space of the local conformal constraints.

\section{Conclusions}\label{sec:Conclusions}

The true value of pure gravity on a torus in 2+1 dimensions is that it is a nontrivial yet completely solvable model that exhibits many of the features of more complicated gravitational systems. It has thus established itself as a valuable testing ground for new gravitational theories such as Shape Dynamics that one can use to learn about the new theory. The main difficulty in constructing Shape Dyamics is to obtain explicit expressions for the Shape Dynamics Hamiltonian on the full ADM phase space, so we are mainly interested in obtaining an explicit Shape Dynamics Hamiltonian. In these investigations we observed
\begin{enumerate}
 \item The explicit (rather than formal) solvability of the initial value problem on the 2-torus (and 2-sphere) in CMC gauge is the technical reason for the explicit constructability of the Shape Dynamics Hamiltonian on these topologies. We find that the Shape Dynamics Hamiltonian formally coincides with the reduced phase space Hamiltonian and that this Hamiltonian is invariant under diffeomorphisms and conformal transformations that preserve the total volume. The difference between the two is however that the Shape Dynamics Hamiltonian is a function of the full ADM phase space that happens to functionally depend only on the image of reduced phase space under the canonical embedding, while the reduced phase space Hamiltonian is a phase space function on reduced phase space itself.
 \item Although we cannot explicitly construct the Shape Dynamics Hamiltonian on higher genus Riemann surfaces, we can construct it perturbatively. In particular we use an expansion that becomes good in a large volume regime. The leading orders of this expansion then turn out to coincide with the temporal gauge Hamiltonian.
 \item The Hamiltonian is in general a nonlocal phase space function. However, it becomes a local phase space function in the large volume limit; in particular one finds that the leading order in a large volume expansion turns the Hamiltonian into the conformal constraint that changes the total volume, so full conformal invariance is attained in this limit.
 \item Since all local constraints are linear in momenta, one can quantize these as vector fields on configuration space. Then gauge invariance implies that the wave function is invariant under the flow generated by these vector fields, which in turn implies that the wave function depends only on reduced configuration space, which is finite dimensional. The Hamiltonian depends only on operators that preserve the reduced phase space, and thus Dirac quantization of the field theory is effectively reduced to reduced phase space quantization.
 \item The construction of Shape Dynamics on the torus relied on the explicit solvability of the modified Lichnerowicz--York equation. This suggests to take the strong gravity BKL limit, where the analogous equation can be explicitly solved as a starting point for quantization, and to introduce the effect of derivative operators through a perturbation series. However it may turn out that alternative quantization strategies are more advantageous as, e.g., the one discussed at the end of section \ref{sec:HigherDimensions}.
\end{enumerate}
Lastly, let us remark that Shape Dynamics is a natural setting to discuss a quantum ``large-CMC-volume/CFT'' correspondence, which we were able to investigate explicitly in an excursion at the end of section \ref{sec:Quantize}, where we found an explicit asymptotic scaling of solutions to the Wheeler--DeWitt equation under conformal transformations, which is very similar to the correspondence explored in \cite{Freidel:2008sh}.

\subsection*{Acknowledgments}

We would like to thank Sean Gryb for discussions. TB acknowledges support by the Netherlands Organisation for Scientific Research (NWO) through a VICI-grant awarded to R. Loll. Research at the Perimeter Institute is supported in part by the Government of Canada through NSERC and by the Province of Ontario through MEDT. This work was funded, in part, by a grant from the Foundational Questions Institute (FQXi) Fund, a donor advised fund of the Silicon Valley Community Foundation on the basis of proposal FQXi-RFP2-08-05 to the Foundational Questions Institute.

\bibliographystyle{utphys}

\end{document}